\documentstyle[12pt]{article}

\begin{document}
\title{Information-theoretic model selection applied to supernovae data}
\vspace{0.6 cm}
\author{ {\sc Marek Biesiada} \\
{\sl Department of Astrophysics and
 Cosmology, } \\
 {\sl University of Silesia}\\
 {\sl Uniwersytecka 4,  40-007 Katowice, Poland}  \\
 biesiada@us.edu.pl}

 \date{}

\maketitle 

\begin{abstract}
\noindent

Current advances in observational cosmology suggest that our
Universe is flat and dominated by dark energy. There are several
different theoretical ideas invoked to explain the dark energy
with relatively little guidance of which one of them might be
right. Therefore the emphasis of ongoing and forthcoming research
in this field shifts from estimating specific parameters of
cosmological model to the model selection.

In this paper we apply information-theoretic model selection
approach based on Akaike criterion as an estimator of
Kullback-Leibler entropy. Although this approach has already been
used by some authors in similar context, this paper provides more
systematic introduction to the Akaike criterion. In particular, we
present the proper way of ranking the competing models based on
Akaike weights (in Bayesian language - posterior probabilities of
the models). This important ingredient is missing in alternative
studies dealing with cosmological applications of Akaike
criterion.

Out of many particular models of dark energy we focus on four:
quintessence, quintessence with time varying equation of state,
brane-world and generalized Chaplygin gas model and test them on
Riess' Gold sample.

As a result we obtain that the best model - in terms of Akaike
Criterion - is the quintessence model. The odds suggest that
although there exist differences in the support given to specific
scenarios by supernova data most of the models considered receive
similar support. The only exception is Chaplygin gas which is
considerably less supported. One can also notice that models
similar in structure i.e. $\Lambda$CDM, quintessence and
quintessence with variable equation of state are closer to each
other in terms of Kullback-Leibler entropy. Models having
different structure i.e. Chaplygin gas or brane-world scenario are
more distant (in Kullback-Leibler sense) from the best one.

\end{abstract}


\newpage

\section{Introduction}

The problem of ``dark energy'' in the Universe is one of the most
important issues in modern cosmology. It appeared after the
discovery of accelerated expansion of the Universe as inferred
from the SNIa Hubble diagram \cite{Perlmutter}. Since then a lot
of specific scenarios have been put forward as an explanation of
this puzzling phenomenon. They fall into two broad categories:
searching an explanation among hypothetical candidates for dark
energy (cosmological constant $\Lambda$ \cite{Perlmutter},
quintessence - evolving scalar fields \cite{Ratra}, Chaplygin gas
\cite{Kam}) or modification of gravity theory (supergravity
\cite{Brax}, brane world scenarios \cite{DGP}).

 In the problem of statistical inference from empirical
data one very often encounters the problem of selecting the best
 approximating model \cite{Burnham}. This is exactly the problem
one has in the context of dark energy where
 there
exist a variety of theoretical ideas of what could be the cause of
accelerating Universe and at the same time there is relatively
little theoretical guidance of which specific model (or a class
thereof) is preferred. Therefore, it is interesting to ask which
cosmological model is the most supported by the data
 which triggered the problem.

In this paper we approach the above mentioned question from the
perspective of information theoretic model selection. One of the
approaches available is that of Akaike initiated by him in early
seventies \cite{Akaike73} and developed during subsequent years
into a simple to use diagnostic called Akaike Information
Criterion (AIC). In cosmology Akaike criterion has first been used
by Liddle \cite{Liddle} and then in papers \cite{Szydlowski}.
Since AIC is relatively unknown among cosmologists while being
being popular, if not standard, in other branches of science (e.g.
biostatistics \cite{Burnham}) Section 2 contains rudimentary
introduction to the ideas underlying AIC closely following
\cite{Burnham}. In Section 3 we briefly outline cosmological
models which are then compared by means of the Akaike criterion.
We illustrate the model selection ideas on Riess' Gold Sample of
supernovae \cite{Riess 2004}. The final section contains results
and conclusions.

\section{Information theoretical model selection criteria}

Akaike information-theoretical model selection criterion is based
on Kullback-Leibler information. Kullback-Leibler information
between two distributions $f(x)$ and $g(x)$ is defined as
\begin{equation}\label{K-L}
    I(f,g) = \int f(x) \ln{ \frac{f(x)}{g(x)}} dx
\end{equation}
The intuitive meaning of $I(f,g)$ (also called K-L divergence) is
the information lost when $g$ is used to approximate $f$. It can
be viewed as an extension of Shannon's entropy and sometimes is
thus referred to as relative entropy. In cosmology, a very
interesting application of this concept has been made by Hosoya et
al. \cite{Hosoya} who proposed the Kullback – Leibler Relative
Information Entropy as a measure of the distinguishability of the
local inhomogeneous mass density field from its spatial average on
arbitrary compact domains.

Let us assume that $f(x)$ denotes the true mechanism behind the
data and $g(x|\theta)$ its approximating model (parametrized by
$\theta$). The problem is that K-L divergence cannot be assessed
without prior knowledge of the true model $f(x)$ as well as
parameters $\theta$ of the approximating model $g(x|\theta)$.
However, given $f(x)$ and $g(x|\theta)$ there exists the ``best''
value of $\theta$ for which Kullback-Leibler divergence is
minimized. The observation that maximum likelihood estimator
$\hat{\theta}$ of $\theta$ parameter is exactly this K-L ``best''
one was the crucial ingredient in Akaike's derivation of his
criterion.

The core result of Akaike was in showing that an approximately
unbiased estimator of K-L divergence is $ln({\cal
L}(\hat{\theta}|data))- K$ where $\cal L$ is the likelihood
function (more precisely its numerical maximum value - taken at
$\hat{\theta}$) and $K$ is the number of estimable parameters
($\theta$s) in approximating model $g(x|\theta)$. For historical
reasons Akaike formulated this result in the following form:
\begin{equation} \label{AIC}
AIC = - 2 ln({\cal L}(\hat{\theta}|data))+ 2K
\end{equation}
which became known as Akaike information criterion. Heuristically
one may think of it as of an estimator of K-L divergence between
the model at hand $g(x|\theta)$ and an unknown true model $f(x)$
which generated the data. In the expression for AIC one can
recognize two terms: the first measuring goodness of model fit (or
more precisely the lack thereof) and the second one (competing)
measuring model complexity (number of free parameters).

Defined in this way AIC value has no meaning by itself for a
single model (simply because the true model $f(x)$ is unknown).
What is useful, instead are the differences $\Delta_i := AIC_i -
AIC_{min}$ calculated over the whole set of alternative candidate
models $i=1,...,N$ where by $AIC_{min}$ we denoted $min\{AIC_i ;
i=1,...,N\}$. Comparing several models, the one which minimizes
AIC could be considered the best. Then the relative strength of
evidence for each model can be calculated as the likelihood of the
model given the data ${\cal L}(g_i|data) \propto
exp(-\frac{1}{2}\Delta_i) $. Relative likelihoods of the models
${\cal L}(g_i|data$) normalized to unity are called Akaike weights
$w_i$. In Bayesian language Akaike weight corresponds to the
posterior probability of a model (under assumption of equal prior
probabilities). The (relative) evidence for the models can also be
judged by the evidence ratios of model pairs $\frac{w_i}{w_j}
=\frac{{\cal L}(g_i|data)}{{\cal L}(g_j|data)}$. If referred to
the best model, evidence ratio gives odds against the given model.
One can easily see (with these definitions) that AIC differences
of 2,4,8,10 correspond to the odds ratios 2.7, 7.4, 54.6 and 148.4
respectively. This justifies the rules of Akaike model selection
that $\Delta_i$ in the range 0 -- 2 mean that model $i$ has almost
the same support from data as the best one, for the range 2 -- 4
this support is considerably less and with $\Delta_i > 10$ model
$i$ is practically irrelevant.

A very similar criterion was derived by Schwarz \cite{Schwarz} in
a Bayesian context. It is known as the so called Bayesian
information criterion (BIC) (\cite{Schwarz}):
\begin{equation} \label{BIC}
BIC = - 2 ln({\cal L}(\hat{\theta}|data))+ K ln(n)
\end{equation}
where $n$ is sample size and as previously $K$ denotes number of
parameters. BIC is not an estimator of K-L divergence -- its
derivation stems from estimating the marginal likelihood of the
data (marginalized over parameters). In cosmological model
selection context BIC was used in \cite{Szydlowski,Szydlo new}.
Its interpretation, however, should go along similar routes as
presented above: values, differences, weights and odds. BIC does
not take the full advantage offered by Bayesian techniques.
Bayesian model averaging approach, although computationally
demanding is by far better. In cosmology it was pursued e.g. by
Kunz, Trotta and Parkinson \cite{Trotta Kunz} (see also references
therein).

It should be noticed that according to some authors
\cite{Parkinson} in the limit of large data (large $n$) AIC tends
to favor models with more parameters while BIC tends to penalize
them.

\section{Cosmological models fitted to supernovae data}

Our aim is to find out what is the degree of support (in terms of AIC)
given by supernovae
data to different cosmological scenarios which might (at least phenomenologically)
describe presently accelerating Universe.

The sample we use is the so called ``Gold'' sample of Riess et al.
\cite{Riess 2004} comprising 157 supernovae compiled from a set of
previously observed SNIa with reduced systematic errors from
differences in calibrations.

To proceeded with fitting the SNIa data we need the
magnitude-redshift relation $m(z,{\cal M}, \theta_{i}) = {\cal M}
+ 5 \log_{10} D_L (z, \theta_i)$ where: by $\theta_i$ we denoted
symbolically cosmological parameters of fitted scenario and $ D_L
(z, \theta_i) = (H_0/c) d_L (z, \theta_i) $ is the luminosity
distance with $H_0$ factored out and the intercept here is defined
as: $ {\cal M} = M - 5 \log_{10} H_0 +25 $ where $M$ is the
absolute magnitude of SNIa. The fitting is performed according to
procedure equivalent to marginalization over the intercept (as
described in \cite{Astier}).

In the framework of Friedman-Robertson-Walker cosmology the
luminosity distance reads:
\begin{equation}
d_L(z) =  (1+z) \frac{c}{H_0} \frac{1}{\sqrt{|\Omega_k|}}
{\cal F} \left( H_0 \sqrt{|\Omega_k|} \int_0^z \frac{d z'}{H(z') } \right) \nonumber\\
 \label{LD}
\end{equation}
where $\Omega_k := -\frac{k}{a_0^2 H_0^2}$ is the curvature term.
The ${\cal F}(u)$ function is defined as ${\cal F}(u) = \sin{u}$
for $k = +1$, ${\cal F}(u) = u$ for $k = 0$ and ${\cal F}(u) =
\sinh{u}$ for $k = -1$.

The estimation of cosmological model parameters was performed
using the maximum likelihood approach. We assumed that supernovae
measurements came with uncorrelated Gaussian errors and in this
case the likelihood function ${\cal L}$ could be determined from
chi-square statistic ${\cal L} \propto \exp{(-\chi^2/2)}$
\cite{Perlmutter}. The $\chi^2$ function here is defined as:
$$ \chi^2 =  \sum_i \frac{(m_i^{th} - m_i^{obs})^2}{\sigma_i^2} $$
where the sum is over the SNIa sample and $\sigma_i$ denote the
(full) statistical error of magnitude determination.

The sections below briefly introduce three types of cosmological
models which will then be compared by using the Akaike criterion.
Formulae therein are given in general form i.e. including the
curvature term. Further on we will restrict our attention to flat
model $k=0$ because the flat FRW geometry is strongly supported by
cosmic microwave background radiation (CMBR) data
\cite{Boomerang}.

\subsection{$\Lambda$CDM model}

Friedman - Robertson - Walker model with non-vanishing
cosmological constant and pressure-less matter including the dark
part of it responsible for flat rotation curves of galaxies (the
co called $\Lambda$CDM model) is a standard reference point in
modern cosmology. Sometimes it is referred to as a concordance
model since it fits rather well to independent data (such like
CMBR data, LSS considerations, supernovae data). In this case the
expansion equation (also called the Friedman equation) reads:
\begin{equation} \label{Hubble_Lambda}
H^2(z) = H^2_0 ( \Omega_m \; (1+z)^3 + \Omega_{\Lambda}  +
\Omega_{k} (1+z)^2)
\end{equation}
The cosmological constant suffers from the fine tuning problem
(being constant, why does it start dominating at the present
epoch?) and from the enormous discrepancy between facts and
expectations (assuming that $\Lambda$ represents
quantum-mechanical energy of the vacuum it should be 55 orders of
magnitude larger than observed \cite{Weinberg}).

\subsection{Quintessence model}

The most popular explanation of the accelerating Universe is to
assume the existence of a negative pressure component called dark
energy. One can heuristically assume that this component is
described by hydrodynamical energy-momentum tensor with $p = w
\rho$ where $-1 < w < -1/3$ \cite{Chiba98}. In such case this
component is called "quintessence".

In quintessential cosmology the Friedman equation reads:
\begin{equation} \label{Hubble_Q}
H^2(z) = H^2_0 ( \Omega_m \; (1+z)^3 + \Omega_Q \; (1+z)^{3(1+w)}
+ \Omega_{k} (1+z)^2)
\end{equation}
where by $\Omega_m$ and $\Omega_Q$ we have denoted present values
of relative contributions of clumped matter and quintessence to
the critical density.

If we think that the quintessence has its origins in the evolving
scalar field, it would be natural to expect that $w$ coefficient
should vary in time, i.e. $w = w(z).$ An arbitrary function $w(z)$
can be Taylor expanded. Then, bearing in mind that both SNIa
surveys or strong gravitational lensing systems are able to probe
the range of small and moderate redshifts it is sufficient to
explore first the linear order of this expansion. Such
possibility, i.e. $w(z) = w_0 + w_1 z$ has been considered in the
literature (e.g. \cite{Weller}). The Friedman equation reads now:
\begin{equation} \label{Hubble_var}
H^2(z) = H^2_0 ( \Omega_m \; (1+z)^3 + \Omega_Q \;
(1+z)^{3(1+w_0-w_1)}\;\exp(3 w_1 z) + \Omega_{k} (1+z)^2 )
\end{equation}

\subsection{Generalized Chaplygin gas cosmology}

In this class of models matter content of the Universe consists of
pressure-less gas
 with energy density $\rho_m$
representing baryonic plus cold dark matter (CDM) and of the
generalized Chaplygin
 gas with the equation  of state
$p_{Ch} =- \frac{A}{{\rho_{Ch}}^{\alpha}}$ with $0\le \alpha \le
1$, representing dark energy responsible for acceleration of the
Universe.

The Friedman equation can be rearranged to the form:
\begin{equation} \label{Hubble_Ch}
H(z)^2 = H_0^2 \left[ \Omega_{m} (1+z)^3 + \Omega_{Ch} \left(A_0 +
(1 - A_0)(1+z)^{3(1+ \alpha)} \right)^{{1\over 1+\alpha}} +
\Omega_{k} (1+z)^2 \right]
\end{equation}
where the quantities $\Omega_i$, $i=m,Ch,k$ represent fractions of
critical density currently contained in energy densities of
respective components.

Generalized Chaplygin gas models have been intensively studied in
the literature \cite{Makler} and in particular they have been
tested against supernovae data (e.g. \cite{BiesiadaGodlowski2005}
and references therein).

\subsection{Brane-world cosmological model}

According to brane-world scenarios \cite{DGP}, our 4-dimensional
Universe is a surface (a brane) embedded into a higher dimensional
bulk space-time in which gravity propagates. As a consequence
there exists a certain cross-over scale $r_c$ above which an
observer will detect higher dimensional effects. Cosmological
models in brane-world scenarios have been widely discussed in the
literature \cite{Jain}. In particular the Friedman's equation
takes here the following form:
\begin{equation} \label{Friedman}
H(z)^2 = H_0^2 \left[ (\sqrt{ \Omega_{m} (1+z)^3 + \Omega_{r_c} }
+ \sqrt{\Omega_{r_c}} )^2 + \Omega_{k} (1+z)^2 \right]
\end{equation}
where: $\Omega_{r_c} = \frac{1}{4 r_c^2 H_0^2}$.
It has been shown in \cite{Jain} that flat brane-world Universe
with $\Omega_m=0.3$ and $r_c = 1.4 \;H_0^{-1}$ is consistent with
current SNIa and CMBR data. Note that in flat (i.e. $k=0.$)
brane-world Universe the following relation is valid:
$\Omega_{r_c} = \frac{1}{4}(1-\Omega_m)^2$.

\section{Results and conclusions}

Table 1 displays the results of fitting the above mentioned models
to the "Gold" sample of SNIa. As already mentioned the flat prior
$k=0$ was assumed. The reason for taking prior assumptions was
that although $\Omega_k$ could have been included as a free
parameter in statistical analysis
(e.g.\cite{BiesiadaGodlowski2005}), the information (even more
precise than achievable this way) about cosmological parameters
like $k$, $H_0$, etc. comes from other types of experiments (such
like CMBR, LSS power spectrum, BBN, gravitational lensing etc.).
In observational cosmology our goal is in building a consistent
picture of the Universe rather than expanding parameter space for
statistical analysis. Matter density $\Omega_m$ was taken as a
free parameter in analysis. Best fit to the supernova data
distinguished an unrealistic value of $\Omega_m = 0.49$ in
quintessential models, which is also reflected in the best fitted
values of $w$ parameters of cosmic equation of state. If one took
a prior on matter density $\Omega_m=0.3$ (as supported by
alternative evidence) one would obtain $w=-1.02 \pm 0.11 $ for
quintessence and $w_0=-1.40\pm 0.25 \;  w_1=1.67 \pm 0.89$ in
models with time varying equation of state, which is similar to
the values from combined evidence (SNIa, LSS, CMBR, lensing)
reported in the literature. However, taking this prior would give
an unfair weighting of models considered so we have assumed matter
density to be a free parameter. The author thanks the referee for
clarifying this point.

A comparison of $\Omega_m=0.3$ prior fit to Chaplygin gas model,
which gives $A_0=0.99 \pm 0.03\; \alpha=1.0 \pm 0.59$, and
respective values from Table 1 are also worth noting. In both
cases the ``correct'' value for $\Omega_m$ is singled out, best
fits for $A_0$ are similar but $\alpha$ fits are drastically
different. The best fitted Chaplygin gas model with $\Omega_m$
prior relaxed is physically equivalent to $\Lambda$CDM while
taking a rigid prior prefers original Chaplygin gas model. This
effect of priors in generalized Chaplygin gas models was also
noted and discussed in \cite{BiesiadaGodlowski2005}.

From Table 2 one can see that the best model - in terms of Akaike
Criterion - is the quintessence model. Therefore this model should
be identified as a reference for calculating Akaike differences,
weights and odds against alternative models considered. The odds
suggest that although there exist differences in the support given
to specific scenarios by supernova data almost all models
considered receive similar support by the data. The support given
to the Chaplygin gas model is considerably less --- odds are
almost 15 to 1 against it when compared to quintessential model.
It is somewhat surprising if one recalls that best fitted
Chaplygin model is phenomenologically equivalent to $\Lambda$CDM
which is the K-L closest one to the quintessential model.

One can also notice that models similar in structure i.e.
$\Lambda$CDM, quintessence and quintessence with variable equation
of state are closer to each other in terms of Kullback-Leibler
entropy. In fact, from purely statistical point of view (apart
from different physical motivations behind each one of them), they
can be considered as a family of nested models. Models having
different structure i.e. brane-world scenario or Chaplygin gas are
more distant (in K-L sense) from the best one. Consequently one
loses more information while fitting these models to supernovae
Gold sample. It should be noticed that in other papers
\cite{Szydlowski} referring to Akaike criterion as a tool for
cosmological model selection authors ignored the issue of odds
against competing model (with respect to the best fitted one) -
treating AIC merely as a tool for ranking based just on numerical
value of AIC. This paper fills this gap in showing how the model
selection procedure should be implemented.

It should be stressed that although it might be tempting to claim
that AIC is equivalent to comparing likelihoold functions (which
is true for models with the same number of parameters fitted to
the data) such altitude is not correct. First of all the
likelihood techniques were developed to estimate parameters of a
given model best fitted to the data. Consequently, the likelihoods
measure goodness of fit not the support for a given model as
compared to competing models. It could intuitively be expected
that these two distinct concepts are connected, and indeed they
are. The proper way to do that is given exactly by the Akaike
criterion, which has a sound theoretical background behind.
Moreover, as demonstrated in this paper AIC results can not be
simply predicted by the number of parameters in the model -- the
best one was found to be quintessence model (with 2 parameters)
closest to it is $\Lambda$CDM (1 parameter) and the next one is
Var Quintessence (3 parameters). They were found to have a similar
support, while the Chaplygin gas model (the same number of
parameters as Var Quintessence) is considerably less supported by
the data.

Table 3 displays analogous information as Table 2 with respect to
the Bayesian Information Criterion (BIC). One can see the
differences in both identifying the best fitted model (now it is
$\Lambda$CDM) and in ranking of remaining ones. The order of
ranking reflects dimensionality of the model --- 1 parameter
models are BIC preferred over 2 or 3 parameter ones (Chaplygin gas
now with 157 to 1 odds against is practically ruled out according
to BIC).

In Table 4 ranking of the models considered according to different
criteria: AIC, BIC and $\chi^2$/d.o.f. are presented. Reasons
behind the ranking for AIC were already revealed: models similar
in structure to the best fitted receive comparable support.
Usually used criterion of $\chi^2$/d.o.f. gives similar ranking as
AIC and BIC criterion apparently penalizes models too much for
over fitting (by giving less support to those with larger number
of free parameters).

One can hope that the future will shed more light on the nature of
dark energy in the Universe. Special surveys (e.g. SNAP) are
designed for this purpose. One should realize however, that the
emphasis of the ongoing and forthcoming research is shifting from
estimating specific parameters of the cosmological model (like the
Hubble constant or deceleration parameter or any other physical
parameter of the theory) to the model selection. Along with
Bayesian techniques (e.g. model averaging) \cite{Trotta Kunz}
information-theoretic model selection approaches are the most
promising for this purpose.

\newpage

{\bf Table 1}\\
Values of best fitted parameters of four models tested.

\vspace{1. cm}

\begin{tabular}{|c|c|}
  \hline
  Model & Best fit model parameters (with $1 \sigma$ ranges) \\
  \hline
  $\Lambda$CDM & $\Omega_m=0.31 \pm 0.04$  \\
Quintessence & $\Omega_m=0.49 \pm 0.06 \; w=-2.40 \pm 1.12 $ \\
Var Quintessence & $\Omega_m=0.48 \pm 0.14 \; w_0=-2.48\pm 1.38 \;  w_1=1.88 \pm 2.59$  \\
  Chaplygin Gas & $\Omega_m=0.31 \pm 0.04 \; A_0=1.00 \pm 0.035 \; \alpha=0.002 \pm 0.088$ \\
  Braneworld & $\Omega_m=0.21 \pm 0.03$   \\

    \hline
\end{tabular}

\vspace{3. cm}

{\bf Table 2}\\
Values of AIC, Akaike differences, Akaike weights $w_i$ (in
Bayesian language equivalent to posterior model probabilities) and
odds against the model (with respect to the best fitted one).

\vspace{1. cm}

\begin{tabular}{|c|c|c|c|c|}
  \hline
  Model &  AIC & $\Delta_i$ & $w_i$ & Odds against \\
  \hline
  $\Lambda$CDM & 179.072 & 1.368 & 0.224 & 1.982 \\
Quintessence &  177.704 & 0. & 0.443 & 1. \\
Var Quintessence & 179.645 & 1.941 & 0.168 & 2.639 \\
  Chaplygin Gas & 183.072 & 5.368 & 0.030 & 14.644 \\
  Braneworld & 180.075 & 2.371 & 0.135 & 3.272 \\

    \hline
\end{tabular}

\newpage

\vspace{1. cm}

{\bf Table 3}\\
Analogous values of Bayesian Information Criterion (BIC), BIC
differences, BIC weights and BIC odds against the model (with
respect to the best fitted one).

\vspace{1. cm}

\begin{tabular}{|c|c|c|c|c|}
  \hline
Model &  BIC & $BIC \Delta_i$ &BIC $w_i$ & BIC Odds against \\
  \hline
 $\Lambda$CDM & 182.128 & 0. & 0.481 & 1. \\
 Quintessence &  183.816 & 1.69 & 0.207 & 2.33 \\
 Var Quintessence & 188.814 & 6.68 & 0.017 & 28.30 \\
 Chaplygin Gas & 192.241 & 10.11 & 0.003 & 157. \\
 Braneworld & 183.131 & 1.00 & 0.292 & 1.65 \\

    \hline
\end{tabular}

\vspace{1. cm}

{\bf Table 4}\\
Ranking of cosmological models fitted to SNIa data according to
AIC, BIC and $\chi^2 / {dof}$ criteria.

 \vspace{1. cm}

\begin{tabular}{|c|c|c|c|}
  \hline
Ranking &  AIC & BIC & $\chi^2$/d.o.f.  \\
  \hline
1. &  Quintessence & $\Lambda$CDM & Quintessence  \\
2. & $\Lambda$CDM  &  Braneworld & Var Quintessence   \\
3. &  Var Quintessence & Quintessence &  $\Lambda$CDM  \\
4. & Braneworld  & Var Quintessence & Braneworld  \\
5. &  Chaplygin Gas & Chaplygin Gas & Chaplygin Gas  \\

    \hline
\end{tabular}

\end{document}